\documentclass[aps,prd,floatfix,reprint,twocolumn,reprint,amsmath,amssymb,superscriptaddress,showpacs]{revtex4-1}
\usepackage{graphicx}
\usepackage{bm}
\usepackage{amsmath}
\usepackage{dcolumn}
\usepackage{dsfont}
\usepackage{amssymb}
\usepackage{tabularx}
\usepackage{array}
\usepackage{float}
\usepackage{color}
\usepackage{epstopdf}
\usepackage{mathrsfs}
\usepackage[colorlinks, linkcolor=blue,anchorcolor=blue,citecolor=blue,urlcolor=blue]{hyperref}

\begin{document}
\title{ Tunable ferroelectricity and anisotropic electric transport in monolayer $\beta$-GeSe }

\author{Shan Guan}
\affiliation{Beijing Key Laboratory of Nanophotonics and Ultrafine Optoelectronic Systems, School of Physics, Beijing Institute of Technology, Beijing 100081, China}
\affiliation{Research Laboratory for Quantum Materials, Singapore University of Technology and Design, Singapore 487372, Singapore}

\author{Chang Liu}
\affiliation{Beijing Key Laboratory of Nanophotonics and Ultrafine Optoelectronic Systems, School of Physics, Beijing Institute of Technology, Beijing 100081, China}
\affiliation{Kunming Institute of Physics, Kunming 650223, China}

\author{Yunhao Lu}
\affiliation{ State Key Laboratory of Silicon Materials, School of Materials Science and Engineering, Zhejiang University, Hangzhou 310027, China}

\author{Yugui Yao}
\email{ygyao@bit.edu.cn}
\affiliation{Beijing Key Laboratory of Nanophotonics and Ultrafine Optoelectronic Systems, School of Physics, Beijing Institute of Technology, Beijing 100081, China}

\author{Shengyuan A. Yang}
\email{shengyuan\_yang@sutd.edu.sg}
\affiliation{Research Laboratory for Quantum Materials, Singapore University of Technology and Design, Singapore 487372, Singapore}

\begin{abstract}
Low-dimensional ferroelectricity has attracted tremendous attention due to its huge potential in device applications. Here, based on first-principles calculations, we predict the existence of spontaneous in-plane electrical polarization and ferroelectricity in monolayer $\beta$-GeSe, a polymorph of GeSe with a boat conformation newly synthesized in experiment. The magnitude of the polarization is about $0.16$ nC/m, which is comparable to that of monolayer SnTe studied in recent experiment, and the intrinsic Curie temperature is estimated to be above 200 K. Interestingly, owing to its puckered structure, the physical properties of $\beta$-GeSe can be easily controlled by strain. The Curie temperature can be raised above room temperature by applying a 1\% tensile strain, and the magnitude of polarization can be largely increased by strains in either armchair or zigzag directions. Furthermore, we find that for the case with electron doping, applying strain can readily tune the anisotropic electric transport with the preferred conducting direction rotated by $90^\circ$, which is connected with a strain-induced Lifshitz transition. The ratio between the effective masses along the two in-plane directions can undergo a dramatic change of two orders of magnitude even by a 2\% strain. Our result reveals monolayer $\beta$-GeSe as a promising platform for exploring ferroelectricity in two-dimensions and for nanoscale mechano-electronic device applications.
\end{abstract}

\maketitle
\section{Introduction}
Achieving ferroelectricity in low-dimensional materials and structures has been a long-sought goal in physics research because of its importance in fundamental science and its great potential for nanoscale device applications~\cite{Setter2006,Scott2007,Nuraje2013}. For conventional three-dimensional bulk ferroelectrics, such as the perovskite BaTiO$_3$ and PbTiO$_3$, the Curie temperature rapidly decreases with the reduction of thickness, due to the increasing cost in energy from the depolarization field, suppressing ferroelectricity in the monolayer limit~\cite{Batra1973,Zhong1994,Dawber2005}. The rise of two-dimensional (2D) materials opens up new perspectives~\cite{CastroNeto2009,Butler2013,Tan2017}. From theoretical calculations, several atomic-thick 2D materials have been predicted to exhibit ferroelectricity, including the distorted 1T-MoS$_2$ monolayer~\cite{Shirodkar2014}, the low-buckled hexagonal III-V binary monolayers~\cite{DiSante2015}, the unzipped graphene oxide monolayer~\cite{Noor-A-Alam2016}, the III$_2$-VI$_3$ van der Waals materials~\cite{Ding2017}, and the monolayer Group-IV monochalcogenides~\cite{Wu2016,Fei2016,Wang2017,Wan2017}. In a recent work, Xiao \emph{et al.} also predicted the first elemental ferroelectric materials in the Group-V monolayers~\cite{Xiao2017}. On the experimental side, Chang \emph{et al.} have successfully detected the ferroelectricity in monolayer SnTe by using the scanning tunneling microscopy and spectroscopy techniques~\cite{Chang2016}. And Zhou \emph{et al.} have detected the ferroelectricity in $\alpha$-In$_2$Se$_3$ nanoflakes by piezo-response force microscopy~\cite{Zhou2017}. These findings stimulate great interest in searching for other 2D ferroelectric materials and in studying the interplay between ferroelectricity and unique properties of 2D materials.

Recently, a new polymorph of GeSe, known as $\beta$-GeSe, has been synthesized in experiment and found to be stable at ambient condition~\cite{Rohr2017}. Its structure has a boat conformation for the Ge-Se six-membered ring, and like $\alpha$-GeSe, it does not possess an inversion center when thinned down to the monolayer limit. Regarding the electronic properties, from calculations, $\beta$-GeSe was shown to be an indirect-gap semiconductor both for its bulk and for its monolayer form~\cite{Rohr2017}. Given that the monolayer $\alpha$-GeSe was previously predicted to be a ferroelectric material~\cite{Fei2016}, one naturally wonders whether monolayer $\beta$-GeSe would also be a 2D ferroelectric. In addition, puckered lattice structure taken by $\beta$-GeSe is especially susceptible to strain tuning. For example, the black phosphorene~\cite{Li2014}, which has the similar structure, exhibit extraordinary properties such as the ultrahigh critical strain ($\sim$30\%)~\cite{Peng2014}, the negative Poisson ratio~\cite{Jiang2014}, the strain-controllable anisotropic transport~\cite{Fei2014} as well as topological properties~\cite{Lu2016}. Since strain-engineering has been proven to be a powerful tool in the field of 2D materials, it is interesting to investigate the strain effects on the possible ferroelectricity and other properties of $\beta$-GeSe.

Motivated by the recent experimental success in the synthesis of $\beta$-GeSe~\cite{Rohr2017} and by the surge of research activities on 2D ferroelectric materials, in this work, through first-principles simulations, we investigate the ferroelectricity and the strain effects in monolayer $\beta$-GeSe. We show that the material has a sizable in-plane spontaneous electric polarization about 0.16 nC/m, which is comparable to that of monolayer SnTe. The intrinsic Curie temperature $T_C$ is estimated to be 212 K by using Monte Carlo (MC) simulations. A rather small strain ($\sim 1\%$) can greatly enhance the value of polarization and raise $T_C$ above the room temperature. Interestingly, we find that strain can also control the anisotropy in electron transport, with the preferred conducting direction rotated by $90^\circ$. With electron doping, the Fermi surface topology changes under strain, corresponding to Lifshitz transitions, and there appears different valley features. Our result reveals monolayer $\beta$-GeSe as a new 2D ferroelectric material and as a promising platform for studying the intriguing interplay between ferroelectricity, lattice strain, electronic transport, and valley degrees of freedom. It may also lead to potential applications in nanoscale ferroelectric and mechano-electronic devices.

\section{COMPUTATIONAL DETAILS}\label{section:methods}
The first-principles calculations were performed based on the density functional theory (DFT), using the projector augmented wave method~\cite{Bloechl1994} as implemented in the Vienna \emph{ab-initio} Simulation Package~\cite{Kresse1993,Kresse1996}. The exchange-correlation functional was modeled within the generalized gradient approximation (GGA) with the Perdew-Burke-Ernzerhof (PBE) realization~\cite{PBE}. A plane-wave basis set with cutoff of 350 eV was used. Monkhorst-Pack $k$-point mesh with size of 18$\times$28$\times$1 was applied for the Brillouin zone (BZ) sampling. The artificial interaction between periodic images was minimized by adding a vacuum layer of 16 \AA\, thickness. The lattice constants and the ionic positions were fully optimized until energy and force are converged with accuracy of $10^{-6}$ eV and 0.005 eV/\AA, respectively. The phonon spectrum calculation was performed with a $7\times7\times1$ supercell using the PHONOPY code through the DFPT approach~\cite{Togo2015}. The spontaneous polarization $P_S$ was calculated using the Berry phase approach~\cite{King-Smith1993,Resta1994}. A supercell of 31$\times$31$\times$1 unit cells was used in the Monte Carlo (MC) simulations. The interaction between nearest neighboring dipole moments in a supercell was described in the mean-field approximation~\cite{Zhong1995,Fei2016}. The effect of van der Waals (vdW) correction was tested (with the DFT-D2 method~\cite{Grimme2006}), which yield qualitatively the same results. Hence, in the following discussion on the monolayer structure, we focus on the result without the vdW correction.

\section{RESULTS}\label{section:results}
\subsection{Ferroelectricity in monolayer $\beta$-GeSe}
Single crystals of $\beta$-GeSe have been experimentally synthesized from $\alpha$-GeSe at high pressure (6 GPa) and high temperature (1200 $^\circ$C), and found to remain stable under ambient conditions~\cite{Rohr2017}. In its monolayer form, $\beta$-GeSe belongs to the space group $Pmn2_1$ with a boat conformation. We have calculated the phonon spectrum for monolayer $\beta$-GeSe (See the Supplemental Material~\cite{SM2017}) and confirmed that the structure is dynamically stable because of the absence of any imaginary frequency in the spectrum.

\begin{figure}[htb!]
\centerline{\includegraphics[width=0.48\textwidth]{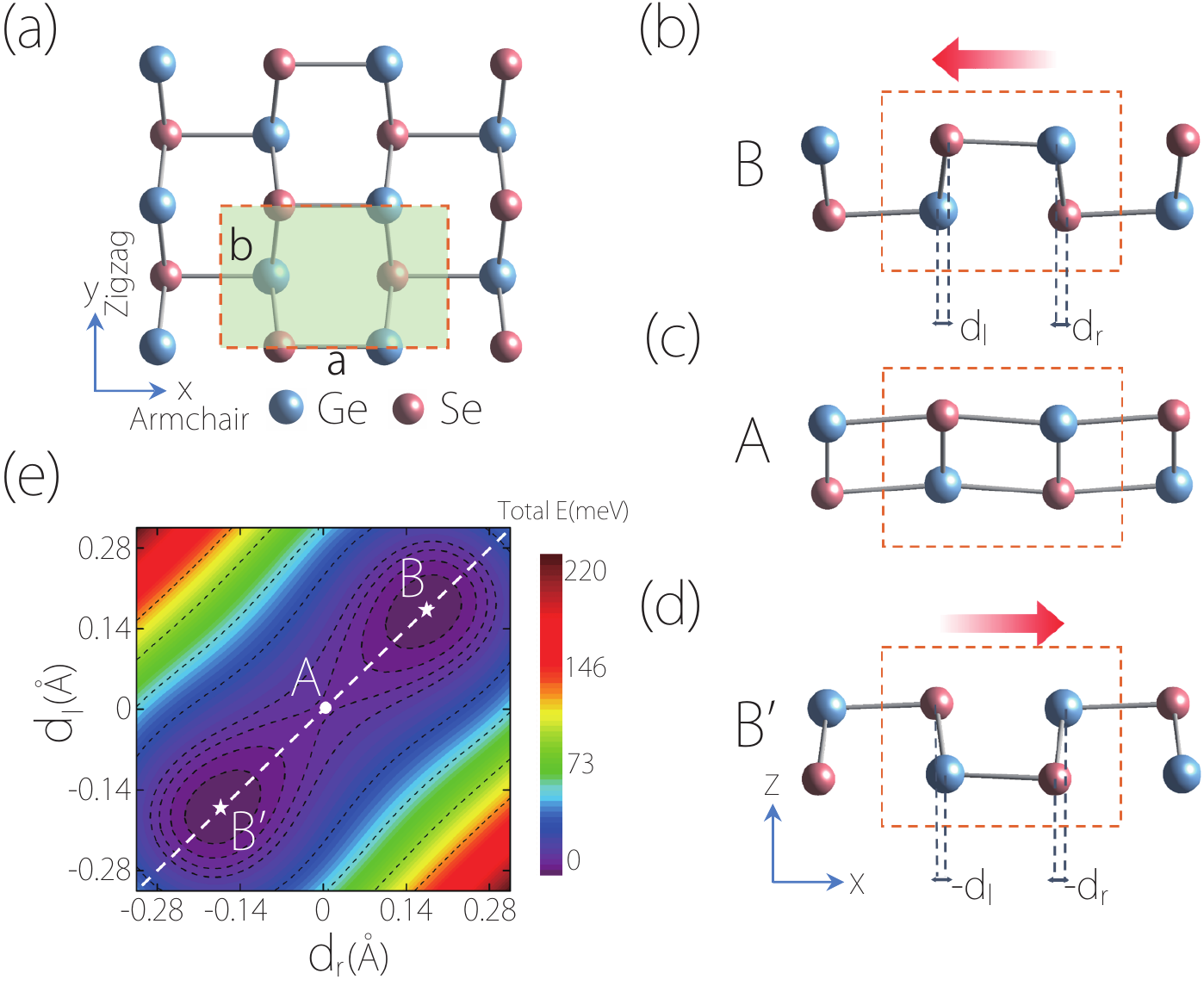}}
\caption{(a) Top view of the crystal structure of monolayer $\beta$-GeSe. The shaded  area marks the unit cell. $a$ and $b$  are the two lattice constants. (b-d) Side views of the three states. State A in (c) is for the undistorted structure with centrosymmetry. B and B' are for the two degenerate ferroelectric structures. They can be considered as distorted from A, during which the centrosymmetry is broken. In (b) and (d), the red arrows indicate the electric polarization direction. $d_\ell$ and $d_r$ are the displacements between the two pairs of neighbouring atoms in the unit cell. (e) Free energy contour plot for monolayer $\beta$-GeSe versus the displacements (d$_\ell$, d$_r$). }\label{fig1}
\end{figure}

The top view of the lattice structure is shown in Fig.~\ref{fig1}(a). The unit cell (indicated by the dashed lines) has a rectangular shape, in which there are two germanium atoms and two selenium atoms. The lattice parameters $a$ and $b$ are calculated to be 5.895 {\AA} and 3.670 {\AA}, which are consistent with the previous work~\cite{Xu2017a}. In our setup as in Fig.~\ref{fig1}(a), the armchair and the zigzag directions are along the $x$ and $y$ axis, respectively. From the side view [Fig.~\ref{fig1}(b)], we can see that the atoms are arranged in the sequence of Ge-Se-Ge-Se in the unit cell, with the Se atoms displaced along the armchair ($x$) direction with respect to the Ge atoms, leading to the breaking of inversion symmetry. The lack of inversion symmetry implies the potential for a spontaneous electric polarization in the material.

To investigate the lattice distortion that breaks the inversion symmetry, we define $d_\ell$ ($d_r$) to be the displacement between the neighbouring Ge and Se atoms along the armchair direction in the left (right) part of the unit cell, as indicated in Fig.~\ref{fig1}(b). When $d_\ell=d_r=0$, we have a centrosymmetric structure, denoted as the A state [see Fig.~\ref{fig1}(c)], which forbids an electric polarization. In comparison, the stable structure in Fig.~\ref{fig1}(b) has $d_\ell=d_r>0$ (denoted as B), and the finite displacements breaks the inversion symmetry. One also notes that via a spatial inversion operation, the B structure is mapped to the B' shown in Fig.~\ref{fig1}(d), which is also stable and degenerate with B in energy. Hence, B and B' belong to a single (ferroelectric) phase.
In Fig.~\ref{fig1}(e), we plot the free-energy contours with respect to the two displacements $d_\ell$ and $d_r$. The result shows that B and B' indeed correspond to the two minimum points on the energy surface. They are connected through a saddle point which is just the centrosymmetric A state. This kind of double well structure is typical for ferroelectric materials, hence its appearance here strongly hints at possible ferroelectricity in monolayer $\beta$-GeSe.

\begin{figure}[!t]
\centerline{\includegraphics[width=0.5\textwidth]{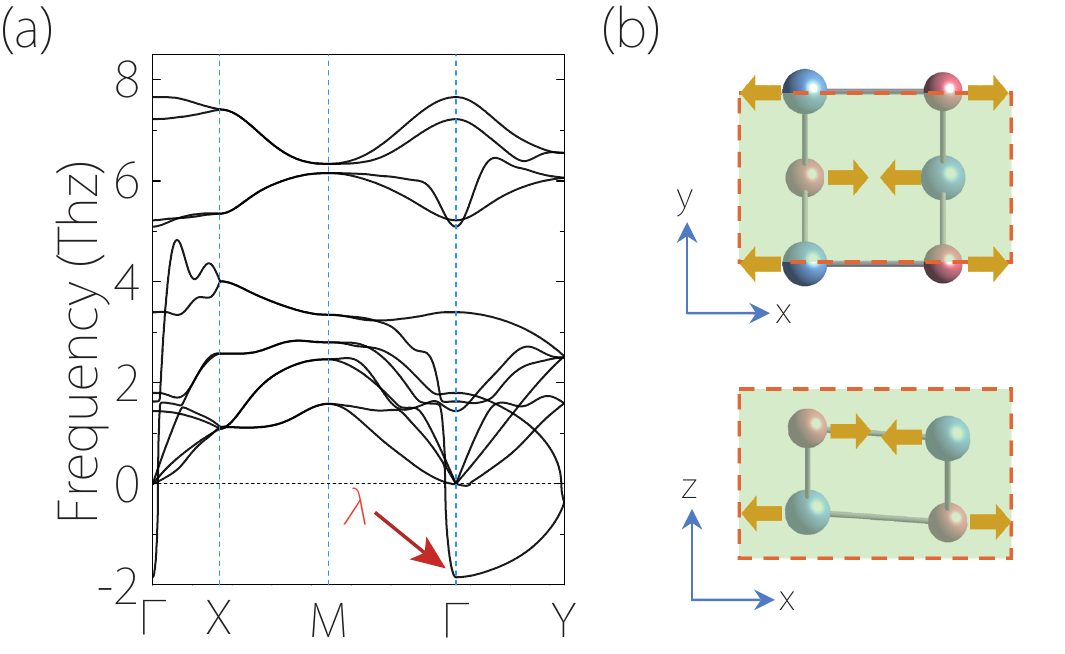}}
\caption{ (a) The calculated phonon spectrum for monolayer $\beta$-GeSe with the \emph{undistorted} centrosymmetric structure [phase A in Fig.~\ref{fig1}(c)]. A pronounced soft optical phonon mode ($\lambda$) at $\Gamma$ can be observed. (b) Schematics showing the top and side views of the phonon eigenvector corresponding to the mode $\lambda$ in (a).}\label{fig2}
\end{figure}

The instability towards lattice distortion for ferroelectric materials is often associated with the soft optical phonon modes for the \emph{undistorted} structure which corresponds to the A state here. In Fig.~\ref{fig2}(a), we plot the calculated phonon spectrum for the A state, in which one observes a pronounced soft optical mode at the BZ center $\Gamma$. From the phonon eigenvector shown in Fig.~\ref{fig2}(b), we find that this mode corresponds to the movement of neighboring atoms of the same atomic plane in opposite directions along $x$ with $d_\ell=d_r$, which just leads to the distortion towards the non-centrosymmetric B or B' structures.

Ferroelectricity in monolayer $\beta$-GeSe is indeed confirmed by our DFT calculation using the Berry phase method~\cite{King-Smith1993,Resta1994}. The spontaneous polarization (per area of the 2D sheet) at zero temperature ($P_S$) is calculated to be 0.159 nC/m, which is comparable to the value for monolayer SnTe ($\sim0.194$ nC/m)~\cite{Wan2017} that was successfully detected in recent experiment~\cite{Chang2016}. The corresponding bulk value is about 0.39 C/m$^2$ if we estimate the effective thickness of the monolayer to be 0.41 nm~\cite{Rohr2017}. The energy difference between the ferroelectric and the centrosymmetric states gives an estimation of the transition barrier $E_B$, which is about 11.66 meV [5.83 meV per formula unit (f.u.)]. The obtained $E_B$ is much smaller than that for the conventional ferroelectric PbTiO$_3$ ($E_B\sim$ 200 meV/f.u.)~\cite{Cohen1992}, indicating that a much lower electric field is required to switch the direction of the electric polarization, which could be advantageous for achieving high energy efficiency for ferroelectric devices. It should be mentioned that a small energy barrier may lead to possible quantum tunneling effects. However, we note that for the confirmed 2D ferroelectric SnTe, the barrier is even smaller, $\sim4.48$ meV/f.u.~\cite{Wan2017}. Yet the ferroelectricity in SnTe is stable up to 270 K, as demonstrated in experiment~\cite{Chang2016}. Therefore, we expect that the ferroelectricity in $\beta$-GeSe can be robust.

\begin{figure}[!t]
\centerline{\includegraphics[width=0.5\textwidth]{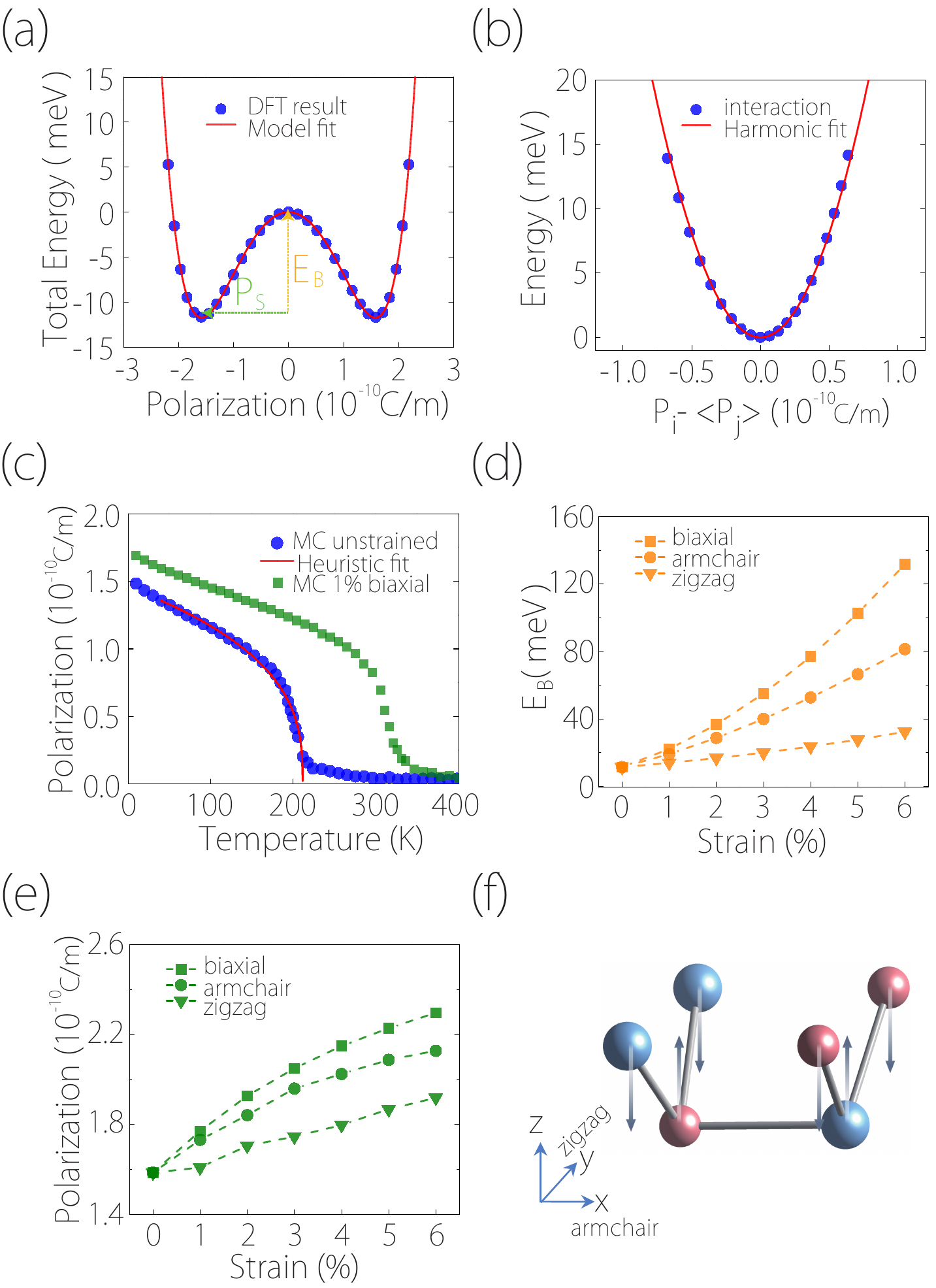}}
\caption{(a) Double-well potential of monolayer $\beta$-GeSe versus polarization. $E_B$ and $P_S$ marks the potential barrier and the spontaneous polarization, respectively. The red line represents the fitting curve of the Landau model [the first three terms in Eq.~(\ref{eq1})]. (b) The dipole-dipole interaction of monolayer $\beta$-GeSe by using mean-field theory. The red line represents a fitting curve by the harmonic approximation. (c) Temperature dependence of the polarization in $\beta$-GeSe obtained by the MC simulations. Here we show the results for the unstrained case and for the $1\%$ biaxially strained case. The red line is a heuristic fit in the vicinity of the Curie temperatures $T_C$ (see the main text). (d) The barrier $E_B$ and (e) the polarization $P_S$ as functions of the applied (three types of) strain. (f) Schematic of the evolution of lattice structure under a tensile strain in the zigzag direction.}\label{fig3}
\end{figure}

\begin{table*}[!htb]
\caption{The ground-state free energy (potential barrier) $E_B$ (meV), the spontaneous polarization $P_S$ (nC/m) at zero temperature,  the fitted parameters A, B, C and D in Eq.~(\ref{eq1}) (their units are chosen such that the free energy is in unit of meV/cell if $P$ is in unit of nC/m), and the estimated Curie temperature $T_C$ (K).}
\label{tab:tab1}
\begin{tabular}{p{2.2cm}<{\centering}|p{2cm} p{2cm} p{2cm} p{2cm} p{2cm} p{2cm} p{2cm} p{2cm}}
  \hline\hline
  Material & $E_B$ & $P_S$ & $A\ (\times10^{2})$ & $B\ (\times10^{4})$ & $C\  (\times10^{6})$ & $D\ (\times10^{2})$ & $T_C$\\ \hline
    $\beta$-GeSe & 11.66 & 0.159 & -15.261 & 1.8389 & 1.7137 & 8.0904 & 212 \\
  \hline\hline
\end{tabular}
\end{table*}

Having confirmed the existence of ferroelectricity in monolayer $\beta$-GeSe, we then check the stability of ferroelectricity by estimating its Curie temperature $T_C$, above which the macroscopic spontaneous polarization is suppressed. Following the approach in Refs.~\cite{Fei2016,Wan2017}, we estimate $T_C$ using the MC method. A Landau-type theory is developed with the polarization $P$ taken as the order parameter. For this purpose, we need to map out the ($d_\ell$, $d_r$) free-energy surface [as in Fig.~\ref{fig1}(e)] as a function of $P$. However, this is a computationally formidable task. Fortunately, one observes the steep gradients traverse to the diagonal line in Fig.~\ref{fig1}(e) (marked by the white dashed line), which indicates that the structure prefers the so-called ``$d$-covariant" states with $d_r=d_\ell$. Therefore, we may consider mapping only the one-dimensional subset of all configurations, leading to a significant reduction of the parameter space. The free energy curve along the $d$-covariant line as a function of $P$ has been calculated and plotted in Fig.~\ref{fig3}(a). Then we can construct a Landau-Ginzburg-type expansion of the total energy in the following form~\cite{Fei2016,Wan2017}:
\begin{equation}\label{eq1}
E=\sum_i\Big[\frac{A}{2}(P_i^2)+\frac{B}{4}(P_i^4)+\frac{C}{6}(P_i^6)\Big]+\frac{D}{2}\sum_{\langle i,j\rangle}(P_i-P_j)^2,
\end{equation}
where $P_i$ is the polarization of the $i$-th unit cell, $\langle i,j\rangle$ indicates the nearest neighbors, and $A$, $B$, $C$, $D$ are constant coefficients. The first three terms capture the anharmonic double-well potential [see Fig.~\ref{fig3}(a)]. The last term describes the dipole-dipole interaction between the nearest neighboring unit cells, which includes the geometry of the system and is crucial for phase transition. In  Fig.~\ref{fig3}(b), we show the comparison between the DFT result based on the supercell calculation and the fit using a quadratic function, which validates the harmonic approximation taken for the interaction term. All the coefficients in Eq.~(\ref{eq1}) are obtained by fitting the DFT results in the mean-field approximation~\cite{Zhong1995,Fei2016}, and their values are listed in Table~\ref{tab:tab1}.

With the effective model in Eq.~(\ref{eq1}), we can proceed to perform the MC simulation to investigate the temperature effects and the ferroelectric phase transition. The simulation results are plotted in Fig.~\ref{fig3}(c). The intrinsic $T_C$ of monolayer $\beta$-GeSe is estimated to be around 212 K. The average polarization $\langle P\rangle$ in the vicinity of the transition follows a heuristic form with $\langle P\rangle=C(T_C-T)^{\delta}$ for $T<T_C$, where $C$ is some constant and $\delta$ is the critical exponent. We fit MC simulation results with the heuristic form, and the obtained $\delta$ is around 0.347,  which is lower than $\delta=0.5$ from the standard mean field theory of the 2D Ising model~\cite{Fridkin2014}. This kind of behavior is consistent with the previous findings in Refs.~\cite{Fei2016,Wan2017} for other 2D ferroelectric materials.

Before proceeding, we mention that in the MC approach, we did not consider the possible coupling term between the polarization and lattice strain, which is a standard approximation and yields good agreement between theory and experiment (e.g., for 2D SnTe, $T_C$ from such MC approach is $\sim 265$ K, very close to the experimental value of $270$ K)~\cite{Wan2017,Chang2016}. Here, to further consolidate our result, we perform the \emph{ab-initio} molecular dynamics simulations, which give a result of $T_C\sim 220$ K (see Supplemental Material~\cite{SM2017}), which agrees well with the MC result.

\begin{figure*}[!htb]
\centerline{\includegraphics[width=0.7\textwidth]{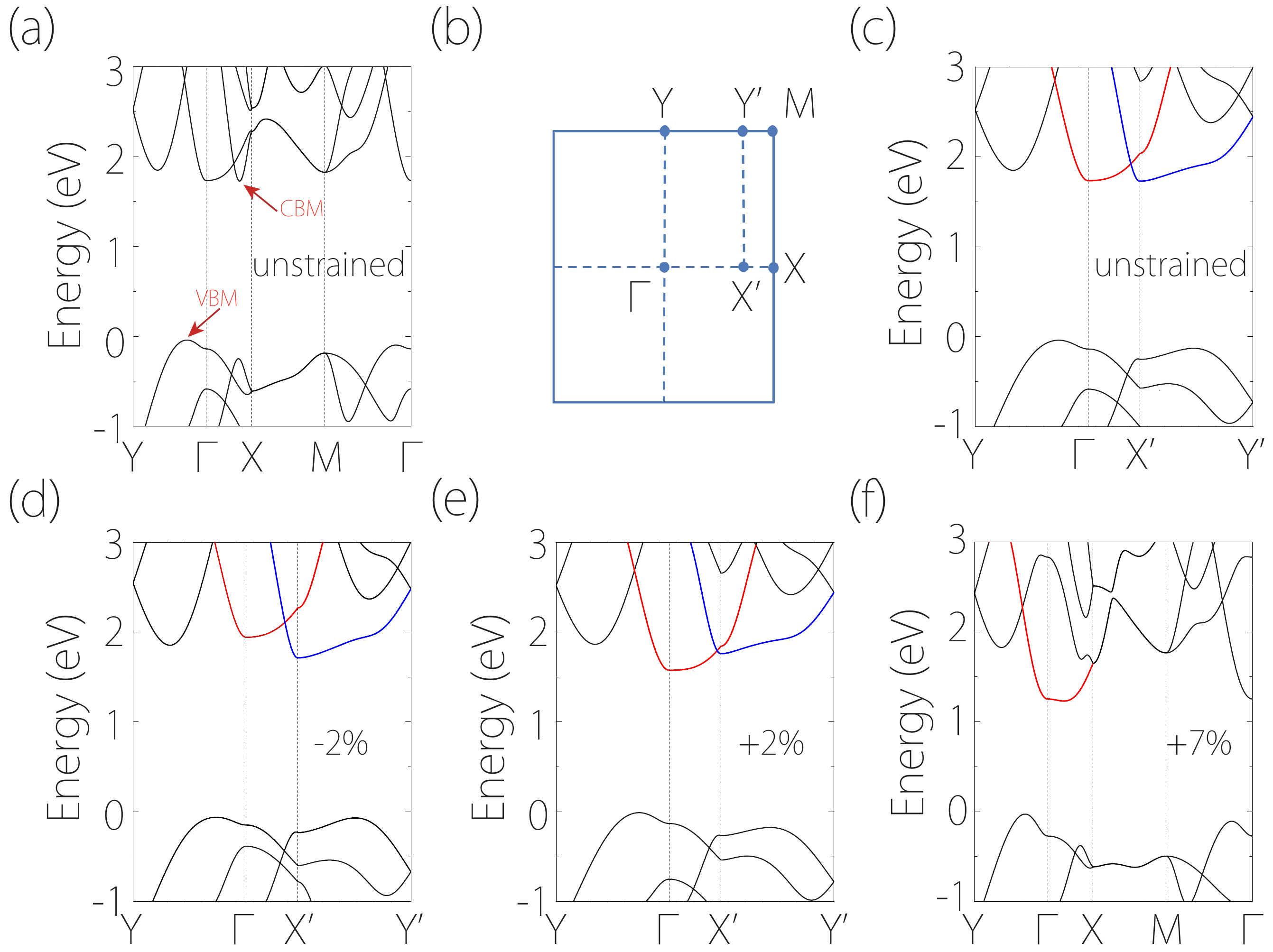}}
\caption{ (a) Electronic band structure for monolayer $\beta$-GeSe without strain. (b) 2D Brillouin zone for monolayer $\beta$-GeSe. X' indicates the location of the CBM for the unstrained case. (c) Plot of the band structure in (a) along a different path, such that the dispersion around the CBM can be clearly shown. (d-e) Band structures for monolayer $\beta$-GeSe (d) with $-$2\% strain, (e) with $+$2\% strain, and (f) with $+$7\% strain. Here, the applied strain is the uniaxial strain along the zigzag direction.}\label{fig4}
\end{figure*}

\subsection{Enhancement of ferroelectricity by strain}
Strain engineering can effectively modify the electronic properties of 2D materials, and even produce remarkable quantum phase transitions, such as the bonding/nonbonding and magnetic phase transition in monolayer MoN$_2$~\cite{Wang2016} as well as the semiconductor/topological-semimetal phase transition in blue phosphorene oxide~\cite{Zhu2016}. Effective strain tuning of ferroelectricity has also been demonstrated on the SrTiO$_3$ thin films, where an enhancement of $T_C$ above room temperature has been observed~\cite{Haeni2004}.

Here we study the effects of three types of strain on the ferroelectric properties. These include the in-plane biaxial strain and the two uniaxial strains along armchair and zigzag directions. In Figs.~\ref{fig3}(d,e), we plot the transition energy barrier $E_B$ and the polarization $P_S$ as functions of the applied strain. One observes that the biaxial and the uniaxial tensile strain along the armchair direction can enlarge both $E_B$ and $P_S$, which can be expected by noticing that both strains tend to increase the relative displacements $d_\ell$ and $d_r$ between Ge and Se atoms. At 6\% biaxial strain, the value of $P_S$ can be increased by more than 40\% to 0.23 nC/m. Interestingly, we find that the the uniaxial tensile strain along the zigzag direction also increases $E_B$ and $P_S$. This seemingly counterintuitive behavior is originated from the puckered lattice structure. As illustrated in Fig.~\ref{fig3}(f), the tensile strain along the zigzag direction tends to pull the ions towards the same $x$-$y$ plane, decreasing the distance between the neighboring atomic sites along the $z$-direction. Then the repulsion between the ions (which tends to maintain the bond length) will lead to increased relative displacement between the neighbors along the $x$-direction, thereby increasing $E_B$ and $P_S$.

The enhancement in $E_B$ and $P_S$ by strain will also increase the Curie temperature. We repeat the MC simulations for the strained monolayer $\beta$-GeSe. The result indeed shows that $T_C$ is increased by the applied strain. Excitingly, the enhancement is quite significant. For example, under a 1\% biaxial strain, $T_C$ can already be increased to about 320 K, above the room temperature [see Fig.~\ref{fig3}(c)]. This implies the great potential for the application of monolayer $\beta$-GeSe in ferroelectric devices, and its sensitive strain dependence also allows promising applications in mechanical sensors.

\begin{figure*}[!htb]
\centerline{\includegraphics[width=0.6\textwidth]{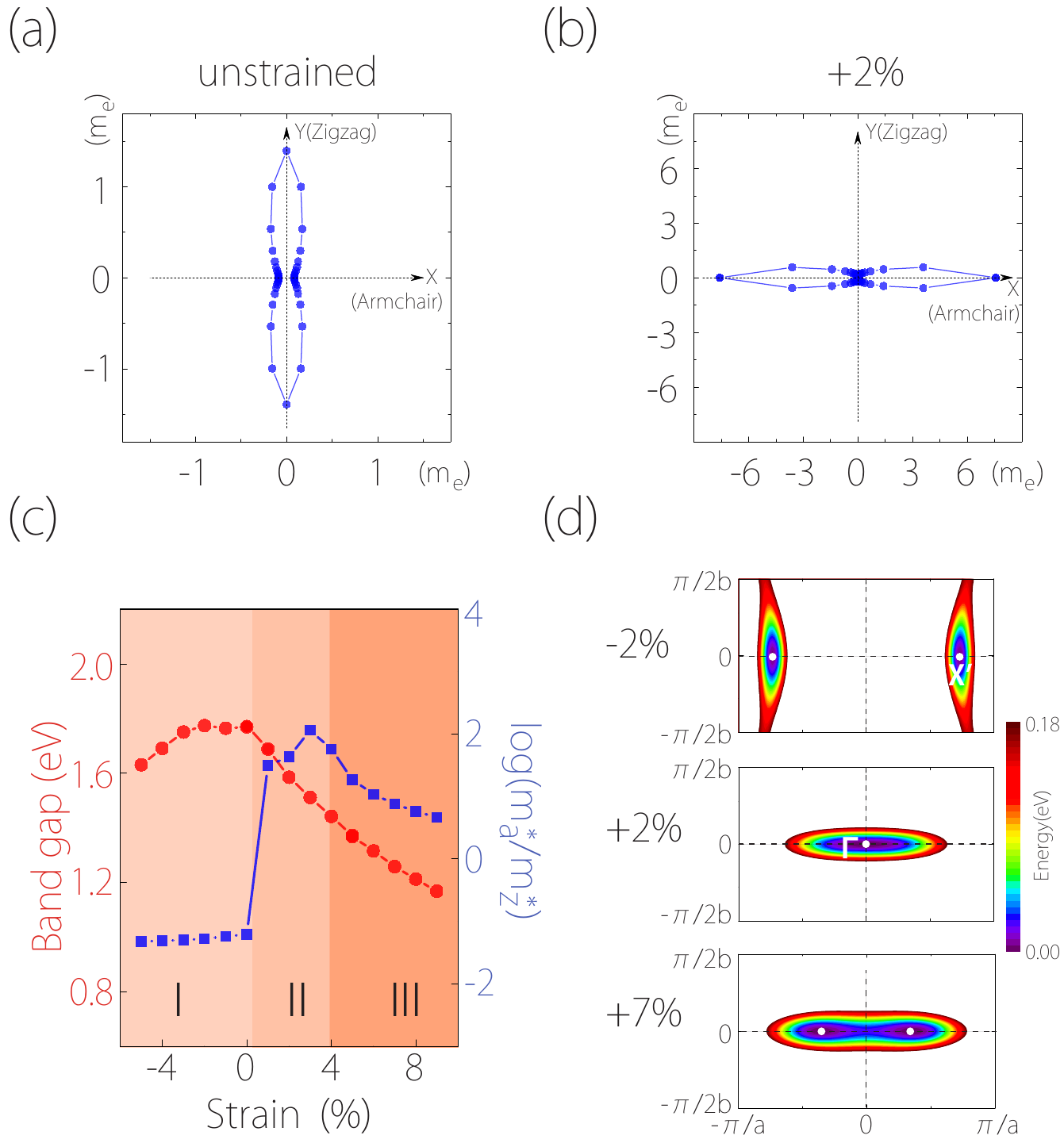}}
\caption{ (a) Electron effective mass of monolayer $\beta$-GeSe according to the spatial direction (a) for unstrained case, and (b) for $+$2\% strained case. (c) The bandgap and the ratio $m_a^*/m_z^*$ (plotted in log scale) plotted as functions of strain. The shaded regions I, II, and III denote the three different cases of the CBM configuration (see the main text). (d) Equi-energy contours for the conduction band under $-$2\%, $+$2\%, and $+$7\% applied strains, corresponding to the three regions in (c). Here, the energy values are referenced to the CBM energy. In these figures, the applied strain is the uniaxial strain along the zigzag direction. }\label{fig5}
\end{figure*}

\subsection{Strain controlled anisotropic transport and Lifshitz transition}\label{3C}

In the following, we focus on the electronic band structure of monolayer $\beta$-GeSe. The DFT result of the band structure is plotted in Fig.~\ref{fig4}(a). One observes that it has a indirect band gap about 1.77 eV with conduction band minimum (CBM) (marked by red arrow) located on the $\Gamma$-X path, whereas the valence band maximum (VBM) is on the $\Gamma$-Y path. As illustrated in Fig.~\ref{fig4}(b), we label the location of CBM by X'. To see the band dispersion around CBM (X' point) more clearly, we re-plot the band structure along the new route in Fig.~\ref{fig4}(c), which includes the dispersion along the perpendicular direction X'-Y'. Interestingly, the band dispersion is highly anisotropic around the CBM (marked by the blue color): the band is highly dispersive along the $\Gamma$-X' direction, whereas its dispersion is quite small along the X'-Y' direction. The $\Gamma$-X' and X'-Y' paths correspond to the armchair and the zigzag directions in real space. The large difference in the dispersions reflects the anisotropy in the lattice structure. As a result, the effective mass of the electron carriers, which are proportional to the inverse of the curvature of the band dispersion, are also highly anisotropic. We plot the electron effective mass along different in-plane directions in Fig.~\ref{fig5}(a), which shows an approximate ``8" shape. The values of the effective mass of electrons differ by an order of magnitude for the two orthogonal directions: the effective mass $m_a^*$ along the armchair direction is about 0.086$m_e$ ($m_e$ is the bare electron mass), whereas the effective mass $m_z^*$ along the zigzag direction is about 1.392$m_e$.

Interestingly, we find that an applied strain can switch the anisotropy in dispersion by 90$^\circ$. In Fig.~\ref{fig4}(c), one observes that there is another conduction band (marked by the red color) with a band minimum located at the $\Gamma$ point, the energy of which is only slightly above CBM. Moreover, the dispersion of this red-colored band is also highly anisotropic: it is highly dispersive along the $\Gamma$-Y (zigzag) direction while nearly flat along the $\Gamma$-X' (armchair) direction. The anisotropy pattern is orthogonal to that of the blue-colored band. (These features have also been confirmed by our calculations using the hybrid functional (HSE06)~\cite{Heyd2003a}. See the Supplemental Material~\cite{SM2017}.) Hence the anisotropic dispersion of the electron carriers will be dramatically changed if the orderings of these two colored bands are switched. In the following, we show that the applied strain can readily generate this band switching and hence change the anisotropy in the carrier dispersion.

In Fig.~\ref{fig4}(d-f), we show the band structure results with applied uniaxial strains along the zigzag direction. One observes that a compressive strain can shift the red-colored band up relative to the blue-colored one, whereas a tensile strain can pull the red-colored band below the blue-colored one. Figure~\ref{fig4}(e) shows that a small (2\%) tensile strain can switch the CBM from X' to $\Gamma$. (The transition occurs at a critical strain less than 1\%.) Comparing Fig.~\ref{fig4}(d) and Fig.~\ref{fig4}(e), one observes that the dispersions of the two bands near their respective minima are almost unchanged. In Fig.~\ref{fig5}(b), we plot the electron effective mass variation for the $+2\%$ strained sample (corresponding to Fig.~\ref{fig4}(e)). One indeed observe that the anisotropy is rotated by 90$^\circ$ compared to the unstrained case: now $m_a^*\simeq 7.576 m_e$ whereas $m_z^*\simeq 0.175 m_e$. If we take the ratio $m_a^*/m_z^*$ as a measure of the anisotropy, then its values changes dramatically from $1/16$ for the intrinsic (unstrained) case to $43$ for the $+2\%$ strained case. The change is above 600 times. In Fig.~\ref{fig5}(c), we plot the ratio along with the variation of the bandgap versus the applied strain. One clearly observes the abrupt jump at the critical strain when the CBM is switched between the two bands (i.e., between I and II).

For the electron doped case, the switch of band ordering discussed above is accompanied by a transformation of the Fermi surface topology, i.e., a Lifshitz transition. And our result shows that there also appear interesting valley features in the conduction band. In Fig.~\ref{fig5}(d), we plot the equi-energy contours around the CBM for three representative strains. One observes that at $-2\%$ strain, there are two separate valleys located on the $\Gamma$-X path, related by the time reversal symmetry. At $+$$2\%$ strain, the CBM is switched to $\Gamma$, hence there is only one valley. In addition, at even larger strains ($+7$\% strain in Fig.~\ref{fig5}(d)), the single valley at $\Gamma$ further split into two along the $\Gamma$-X path, corresponding to the band structure in Fig.~\ref{fig4}(f). The three cases in Fig.~\ref{fig5}(d) correspond to the three different shaded regions in Fig.~\ref{fig5}(c).

In the above discussion, we focus on the uniaxial strain along the zigzag direction. These effects can also be achieved by biaxial strain or uniaxial strain along the armchair direction. We find that the results for the armchair uniaxial strain and the biaxial strain are similar to that for the zigzag uniaxial strain (see Supplemental Material~\cite{SM2017}).

\begin{figure}[!htb]
\centerline{\includegraphics[width=0.5\textwidth]{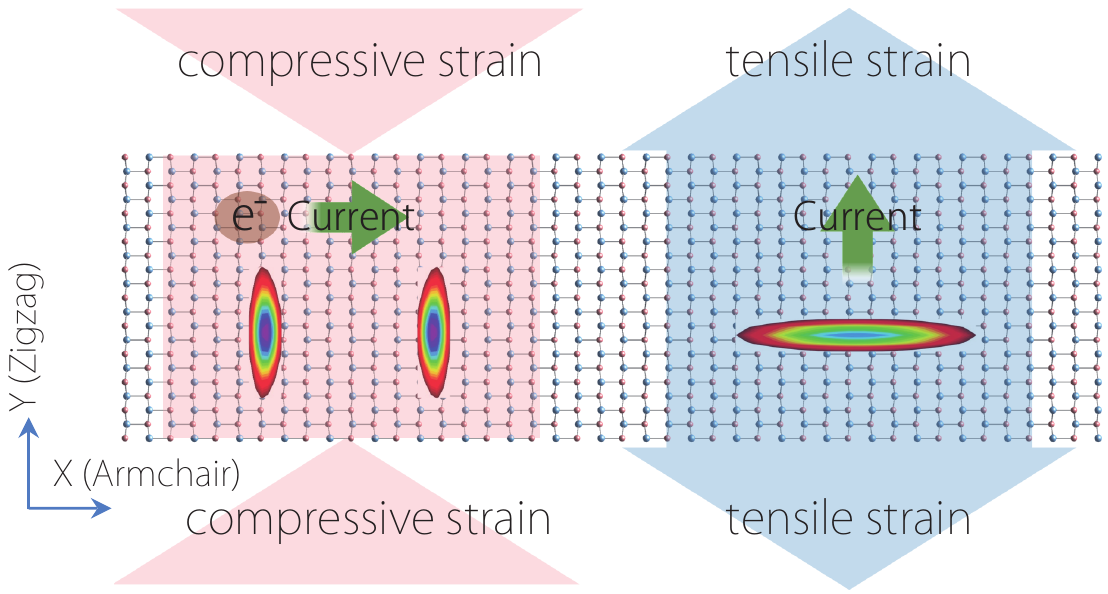}}
\caption{Schematic of strain-induced rotation of electrical conduction anisotropy in monolayer $\beta$-GeSe.}\label{fig6}
\end{figure}

The strong anisotropy in the band dispersion will be reflected in electric transport. For example, the small $m_a^*/m_z^*$ ratio for the unstrained or compressed case indicates that the electric conduction (conductivity) is easier (larger) in the armchair direction. Across the transition from Region I to Region II in Fig.~\ref{fig5}(c), the easy-conduction direction is switched to the zigzag direction. These two scenarios are schematically shown in Fig.~\ref{fig6}. The sudden change induced by strain can be directly probed in the electric transport measurement. In addition, the abrupt change in Fermi surface topology discussed above can be probed as the change in magnetic quantum oscillations (like the Shubnikov-de Haas oscillations) under strain.

\section{Discussion and conclusion}

As we have mentioned, the 3D bulk $\beta$-GeSe have already been synthesized in experiment and is stable under ambient conditions~\cite{Rohr2017}. This would greatly facilitate the realization of the 2D monolayer, e.g., by using the mechanical exfoliation method. Experimentally, it has been demonstrated that the structurally similar Group-IV monochalcogenides can be thinned down to few layers by mechanical exfoliation~\cite{Ye2017}. Therefore, we expect that monolayer $\beta$-GeSe could also be fabricated in the near future. Once realized, the ferroelectricity can be detected by the standard probes, such as the synchrotron X-ray scattering~\cite{Fong2004}, ultraviolet Raman spectroscopy~\cite{Tenne2009}, polarized second-harmonic generation measurement~\cite{Sheng2010}, piezoresponse force microscope~\cite{Garcia2009}, and STM/STS measurements~\cite{Chang2016}.

Electronic and optical properties of $\beta$-GeSe were also theoretically studied in Ref. 37, in which the strong anisotropy in carrier transport and the strain effects on the bandgap where reported. In comparison, the present work reveals that strain can tune the ferroelectricity to a large extent, switch the anisotropy in carrier transport, and induce interesting Lifshiftz transition for doped samples, associated with changes in the valley configurations. These points have not been addressed before.

Techniques for applying strain on 2D materials have been well developed. For example, strain can be applied by using the beam bending apparatus~\cite{Conley2013} or by using the AFM tips~\cite{Lee2008}.  Generally, tensile strains would be easier to apply than compressive strains, because 2D materials tend to buckle under compression. Experimentally, small compressive strains may be achieved by attaching the 2D sample to substrate and applying compression on the substrate.

2D materials typically have excellent mechanical flexibility. For example, graphene, MoS$_2$, and black phosphorene can sustain strains above 25\%~\cite{Lee2008,Castellanos-Gomez2012,Bertolazzi2011,Peng2014}. Experimentally, strains above 15\% have been demonstrated~\cite{Lee2008,Kim2009}. For monolayer $\beta$-GeSe, our estimation shows that its critical tensile stain can be up to 16\% (see Supplemental Material~\cite{SM2017}). We have calculated the Young¡¯s modulus, which is about 83 GPa along the zigzag direction and about 85 GPa along the armchair direction. These values are comparable to that of phosphorene ($\sim166$ GPa along the zigzag direction and $\sim44$ GPa along the armchair direction)~\cite{Wei2014}, and is less than that for graphene ($\sim1000$ GPa)~\cite{Lee2008} and MoS$_2$ ($\sim330$ GPa)~\cite{Castellanos-Gomez2012}. These results indicate the feasibility of the strain engineering on $\beta$-GeSe discussed in this work.

The major band structure features discussed in Sec.~\ref{3C} have been checked by our calculations including SOC or using hybrid functional method (see Supplemental Material~\cite{SM2017}). Hence our predictions are robust. The band-switching transition occurs at a very small strain around $1\%$. Similar strain effect on anisotropic transport was also predicted in monolayer black phosphorene~\cite{Fei2014}. We mention that, for monolayer $\beta$-GeSe, the required strain for the switching is much smaller than that in monolayer black phosphorene ($\sim 6\%$)~\cite{Fei2014}. This could make the effect for monolayer $\beta$-GeSe more easily observed in experiment.

In conclusion, based on first-principles calculations, we have demonstrated ferroelectricity in monolayer $\beta$-GeSe. We find that the spontaneous polarization and the transition energy barrier can be significantly enhanced by various types of strains. The Curie temperature is estimated to be 212 K, and it can be increased above room temperature by applying a $\sim1\%$ tensile strain. From band structure analysis, we identify highly anisotropic dispersions for the electron carriers. We find that the preferred conduction direction can be rotated by $90^\circ$ under strain. Particularly, the ratio between the effective masses along the two in-plane directions can undergo a abrupt change of two orders of magnitude by a 2\% strain. For electron doped case, the effect is associated with a strain-induced Lifshitz transition and there also appear interesting valley features in the band structure. Our findings thus provide a promising platform to explore the intriguing physics of ferroelectricity in two dimensions and reveal the great potential of $\beta$-GeSe for the nanoscale mechano-electronic device applications.

\begin{acknowledgments}
The authors thank D.L. Deng for helpful discussions. The work is supported by the MOST Project of China (Grants No. 2014CB920903), the NSF of China
(Grants Nos.11574029, 11734003), the National Key R\&D Program of China (Grant No.2016YFA0300600), the Singapore Ministry of Education Academic Research Fund Tier 1 (SUTD-T1-2015004). We acknowledge computational support from the Texas Advanced Computing Center and the National Supercomputing Centre Singapore.

\end{acknowledgments}

%

\end{document}